# ROI Extraction in Thermographic Breast Images Using Genetic Algorithms

L. C. Mendes[1], E. O. Rodrigues[2], S. C. Izidoro[3], A. Conci[4] and P. Liatsis[5]
[1,3]Institute of Technological Sciences, Universidade Federal de Itajubá, Itabira, Minas Gerais, Brazil
[2]Academic Department of Informatics, Universidade Tecnológica Federal do Paraná, Pato Branco, Parana - Brazil
[4]Computer Science Department, Universidade Federal Fluminense, Niteroi - Rio de Janeiro, Brazil
[5]Department of Computer Science and Electrical Engineering, Khalifa University of Science and Technology, Abu Dhabi - UAE
*lucasmendes@unifei.edu.br, erickrodrigues@utfpr.edu.br, sandroizidoro@unifei.edu.br, aconci@ic.uff.br, panos.liatsis@ku.ac.ae*

***Abstract*—This work proposes the use of Genetic Algorithms (GA) to identify the area of the breast from the background in thermographic breast images. The proposed method uses color information, a fitness function based on cardioids, and GA. This is the first work in the literature to propose a Region of Interest (ROI) extraction based on GA and cariods. ROI extraction can improve the accuracy of cancer detection and assist with the standardization of acquisition protocols. The method is able to successfully separate the breast region in 52 out of 58 images, while being fully automatic, and not requiring manual selection of seed points.**

## I. Introduction

The demand for medical diagnosis systems has increased over recent years, together with the requirement for faster and more accurate analysis. Thermographic imaging is widely used in the detection of breast cancer [1]–[5], with one of its key parts being Region of Interest extraction. Appropriate ROI extraction can improve the odds of a medical diagnosis system correctly detecting an image with the presence of cancer, as it excludes non-related information.

Breast cancer is the most frequent cancer among women, impacting on 2.1 million women each year [6]. The majority of diagnostic tools tend to be expensive, however, thermographic imaging has been identified as a potential and robust screening tool for early cancer detection [7], [8].

Evolutionary algorithms (EA) or Genetic algorithms (GA) are a branch of metaheuristics [9], based on natural selection and evolution. GA provide a useful tool in problem areas that do not yield readily to standard approaches [10]. GA is a stochastic algorithm that outputs a near-optimal solution as long as the algorithm runs for sufficient times. Although it relies on natural stochastic mechanisms, its algorithmic pattern is entirely reproducible when information about the seed used to generate random numbers is preserved.

To the best of the authors' knowledge, this work is the first in the literature to propose the use of cardioids and GA for segmentation of the breast region. In what follows, Section II provides a brief literature review, Section III describes the proposed approach, and Sections IV and VI provide the obtained results and conclusions, respectively. Section V provides a brief discussion.

## II. Literature Review

Region of interest (ROI) extraction [11] is an important preprocessing step when it comes to image analysis and classification. Background information usually reduces the accuracy of classifiers, as potentially biased, unwanted information is introduced to the context of the problem [12]. Figure 1 highlights the basic steps of a common image-based classification framework [13].

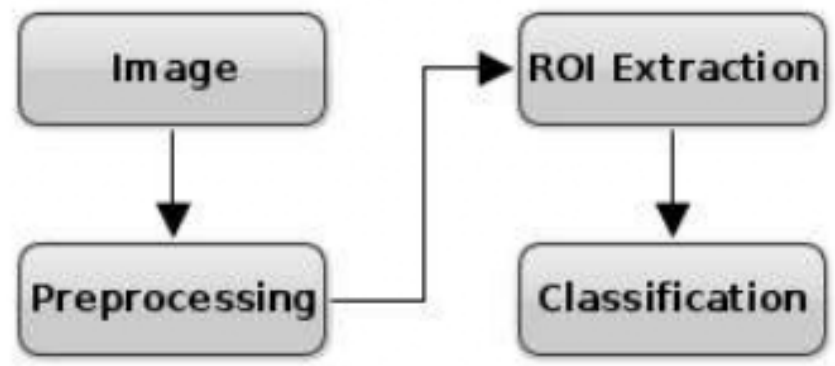

Fig. 1. Fundamental steps in Image-Based Classification.

ROI extraction can be viewed as a pre-processing step or as a separate step, normally applied following image pre-processing. For instance, Suganthy et al., [14] apply anisotropic diffusion to reduce noise before ROI extraction.

Irrespective of its place in the general methodology, ROI extraction is a key step for several image processing approaches such as traffic sign detection [15], thermographic breast image classification [8], finger-knuckle prints [11], computed tomography [16], palm print recognition [17] and many others [12], [18]–[20].

In a previous work [21], we used genetic algorithms to delineate the pericardium contour around the human heart, improving the classification results and the separation of the epicardial and mediastinal adipose tissues. In this work, we use the same principle while aiming to separate the region of interest (i.e., breast) in infrared images, excluding the background and other irrelevant structures.

Suganthi et al., [14] proposed a ROI extraction using anisotropic diffusion and level set-based snakes. Although their method provides good results, it requires some manual segmentation of the region or selection of seed points. In contrast, the algorithm proposed in this work does not require any seed points and is able to adapt to healthy and unhealthy breast images.

## III. Materials and methods

This section describes the proposed methodology, including a brief description of GA, the parameters used to validate the algorithm, provides an overview of the dataset and the details of the cardioid approach.

### A. Dataset

We use infrared images of the breast from the Database for Mammography Research with Infrared Images (DMR-IR) [7], which can be found at *http://www.visual.ic.uff.br/dmi*. These images were captured using the FLIR ThermaCam S45 camera. The project was approved by the Ethical committee of the Federal University of Pernambuco. Thermograms were recorded at the University Hospital and registered with the Brazilian Ministry of Health.

These images were taken from patients and volunteers older than 35 years and corrected for relative humidity and temperature of the room. Patients were requested to wait for 10min in order to stabilize their metabolism. A total of 58 images were randomly selected for this work. The dataset contains images of healthy and sick breasts. Sick breasts include breast diseases such as cancer.

### B. Genetic Algorithm

GA exhibit a unique characteristic, which differs from traditional search methods, i.e., they explore the search space by considering different sets of solutions (individuals) in parallel [22]. This concept is what makes GA a good option in solving this problem as images contain a total of 307200 pixels (480x640), which enables parallel evaluation of cardioids. A cardioid is heart-shaped shape, described in cartesian form as:

$$x = r\cos\theta \tag{1}$$

$$y = r\sin\theta \tag{2}$$

However, it is more convenient to use the polar coordinates form for the cardioid, given by:

$$r = a(1 + \sin\theta) \tag{3}$$

where $a$ is the base of the cardioid, which is used to define the size of the cardioid in each generation of the GA. $r$ describes the radius of the cardioid in relation to its center for each angle $\theta$, varying from 0 to $2\pi$.

As thermographic images are displayed as greyscale images, we separate the 8 bit interval [0,255] in four groups, i.e., $A = [0, 44]$, $B = [45, 119]$, $C = [120, 209]$ and $D = [210, 255]$, as shown in Figure 2. Groups A and D (higher temperatures) should not be engulfed by the cardioid. Group A represents the background of the image and Group D contains areas of the body, which are irrelevant in terms of breast analysis, typically corresponding to the belly and the arms. Group C typically contains the area of interest, while Group B can be found in the arms, abdomen, and chest.

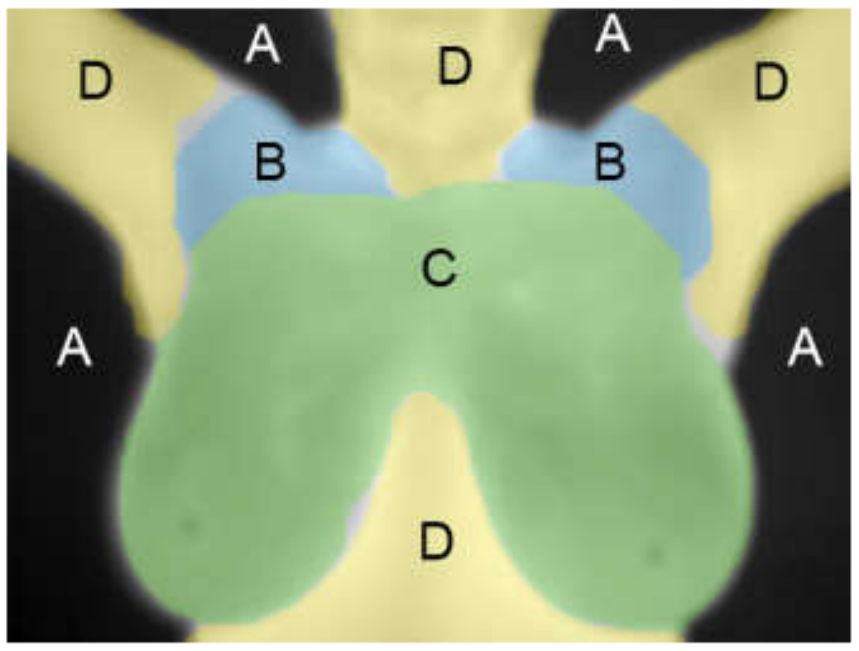

Fig. 2. Grayscale groups used in the fitness function. Pixel group C is indicative of the area of interest.

The fitness function of the GA is described by Equation 4, where the weights (W) were empirically assigned to each group (A to D) based on its importance. Specifically, the values of the weights are as follows $W_A = 300$, $W_B = 50$, $W_C = 250$, $W_D = 275$ and $W_T = W_A + W_B + W_C + W_D$, and the fitness function $f$ is given by:

$$f = \frac{(BW_B + CW_C) - (AW_A + DW_D)}{|W_T|} \tag{4}$$

Each individual in the GA population is composed of three gray code vectors (which is also a binary code). The first part describes the cardioid's position in the x axis, the second part describes its position in the y axis, and the last part describes the cardioid base. Caruana and Schaffer [23] argue that gray coding is generally superior to binary coding for function optimization using GA. Analysis suggests that gray coding eliminates the "Hamming cliff" problem, which creates difficult transitions when using a binary representation.

We start with 60 randomly generated individuals in the population. Next, the algorithm starts a loop of $n$ generations. We use the tournament selection method [24], which consists of randomly choosing $k$ individuals in the population and picking the one with the higher fitness. In our case, $k$ was 2. This operation is repeated twice and both individuals selected by the tournament have 70% chance to be subjected to crossover.

The crossover of the individuals $(i_1, i_2)$ generates 2 children $(i_{c1}, i_{c2})$ that aggregate 2 pieces of each vector of each individual. The vectors receive a random "cut" position and the child $i_{c1}$ is assembled from the first part of $i_1$ and the second part of $i_2$, while $i_{c2}$ is assembled using the first part of $i_2$ and the last part of $i_1$. This is called the one-point crossover, i.e., when a single cut in the genotype is performed [25].

After that, both children are subject to a 2% probability of mutation. In case a child enters the mutation process, each vector has a 50% chance of going through a mutation. The one-point mutation was used in all the phases.

After 25 iterations (i.e., half the size of the population), all the new individuals compose a new population. If this new population does not have 60 individuals, then all of the individuals with the higher fitness scores are transferred to the new population, and the individual with the highest score always goes to the new population (elitist strategy).

The GA evaluates and organizes individuals by their fitness function. Function 4 analyses each pixel of the image and evaluates whether the pixel radius, in relation to the center of the cardioid, is equal or less than the individual cardioid's. The pixels inside the cardioid are sorted into 4 groups based on their grayscale value, as previous addressed. Algorithm 1 illustrates the GA in the form of a pseudo-algorithm.

**Data:** $\Omega$ is the population (60 individuals), *n* is the total amount of generations of the algorithm, *i* is the individual *i* = [x position, y position, base], $\mu$ represents the mutation operation, $\Phi$ represents the cross-over operation and *random(0,100)* returns a randomly generated number between 0 and 100.

```
begin
    Create a random population with 60 individuals and
      place it in Ω;
    while iterations < n do
        Select i1 and i2 in Ω;
        if random(0,100) ≤ 70% then
            ic1, ic2 = Φ(i1, i2)
        if random(0,100) ≤ 2% then
            if random(0,100) ≥ 50% then
                μ(ic1)
            if random(0,100) ≥ 50% then
                μ(ic2)
        iterations++;
        Place the best individual in Ω and remove the
          worst ones (the population has a maximum
          number of 60 individuals);
    end
    Choose the best individual in Ω;
end
```

**Algorithm 1:** Proposed genetic algorithm.

## IV. Results

Figure 3 shows the top individuals (the ones with the highest fitness score) in four healthy women after the convergence of the GA. Figure 4 shows the top individuals in four women affected by breast disease. The parameters of the GA were: (1) population size: 60 individuals, (2) number of generations: 50, (3) crossover rate: 70%, (4) mutation rate: 2%, (5) tournament: 2.

In order to verify the accuracy of our findings, the GA was run 30 times and the average, minimum, maximum, mean, and median fitness values were recorded for 10, 50, 100, and 200 generations. This was supported parameter optimization, while avoiding unnecessary use of computing resources.

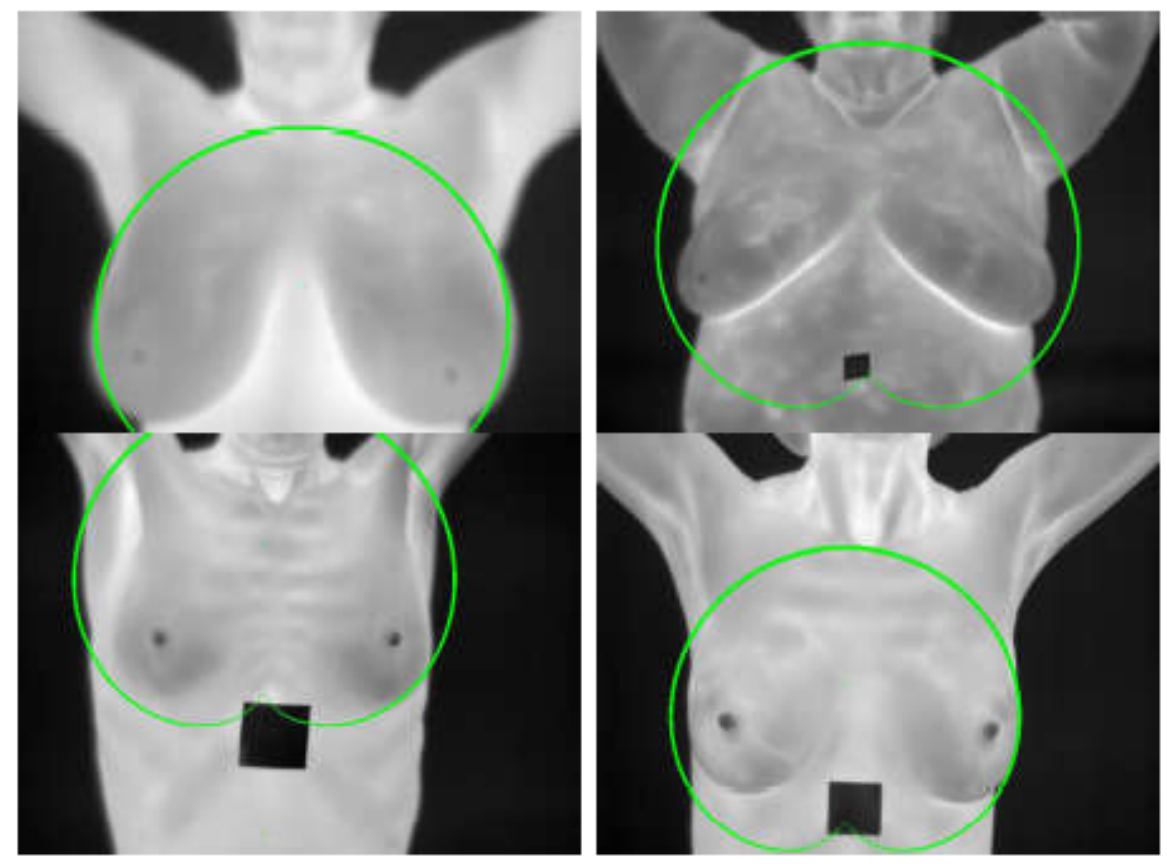

Fig. 3. Examples of ROI detection in healthy women.The ROI is the area within the green curve.

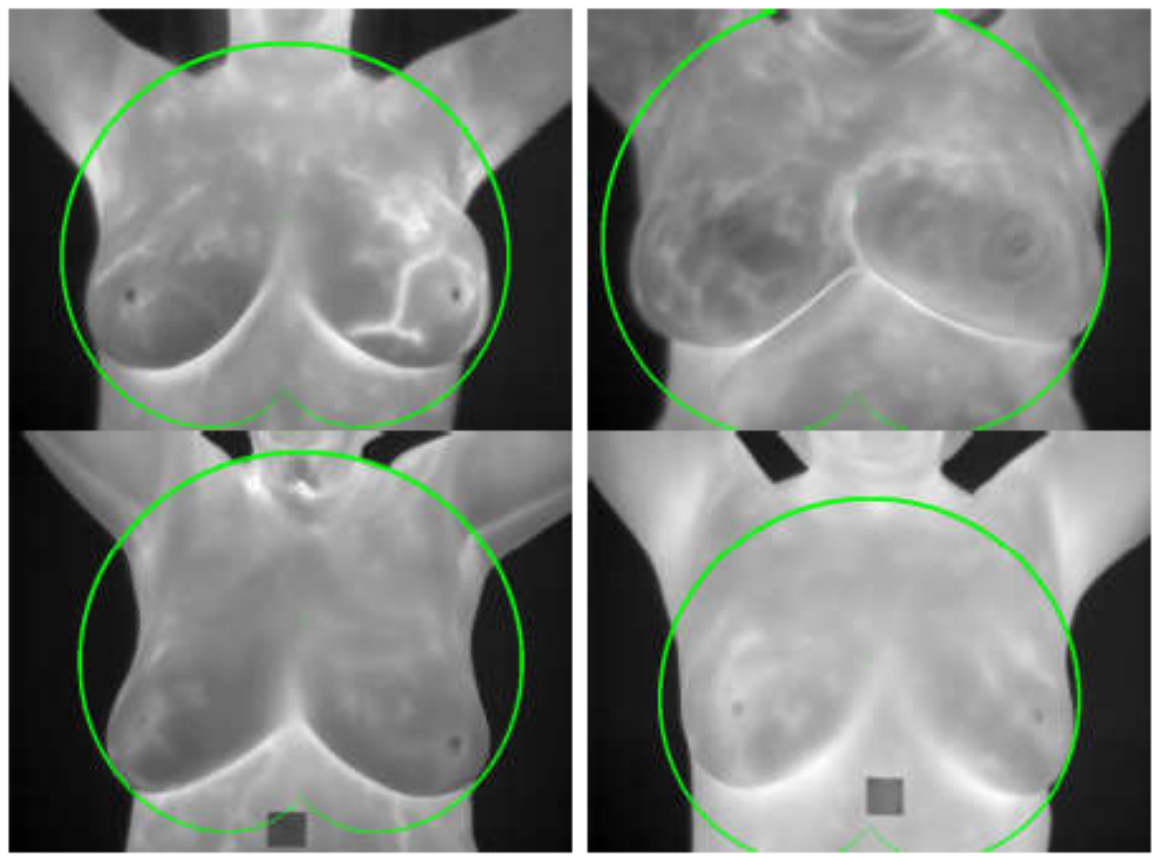

Fig. 4. Examples of ROI detection in women with breast disease.The ROI is the area within the green curve.

TABLE I
FITNESS STATISTICS FOR VARYING GENERATION NUMBERS

| Generations | 10 | 50 | 100 | 200 | SD |
|---|---|---|---|---|---|
| **Mean** | 3.75 | 5.40 | 5.75 | 5.72 | 0.94 |
| **Median** | 3.81 | 5.58 | 5.58 | 5.88 | 0.94 |
| **Min** | 3.76 | 5.30 | 5.82 | 5.40 | 0.90 |
| **Max** | 3.90 | 5.58 | 5.82 | 5.88 | 0.93 |
| **Time (s)** | 407.78 | 1,870.47 | 4,778.10 | 8,691.61 | - |

From Table IV, it is possible to observe a substantial increase in the average fitness score from 10 to 50 generations (54.55%), and a slower increase right after, between 50 to 200 generations (4.08% from 50 to 100 and 2.55% from 100 to 200). Meanwhile, the time increases nearly linearly with the number of generations. In summary, a sharp evolutionary pressure is observed in the first 50 generations. Following this, there is little variation in terms of fitness score statistics.

Figure 5 shows the best cardioids created over the generations. Figure 5-(a) shows a healthy case and Figure 5-(b) a sick case. Color red represents 10 generations, blue 50 generations, yellow 100 generations and green 200 generations. In Figure 5-(a), yellow and green cardioids overlap.

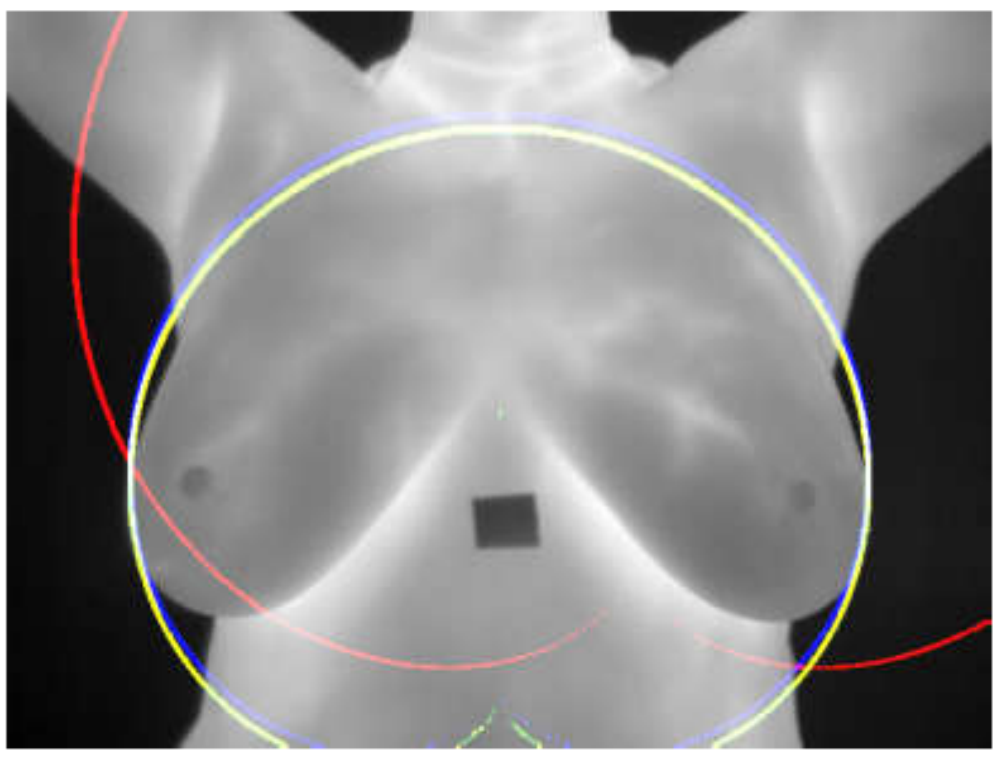

(a) Healthy case.

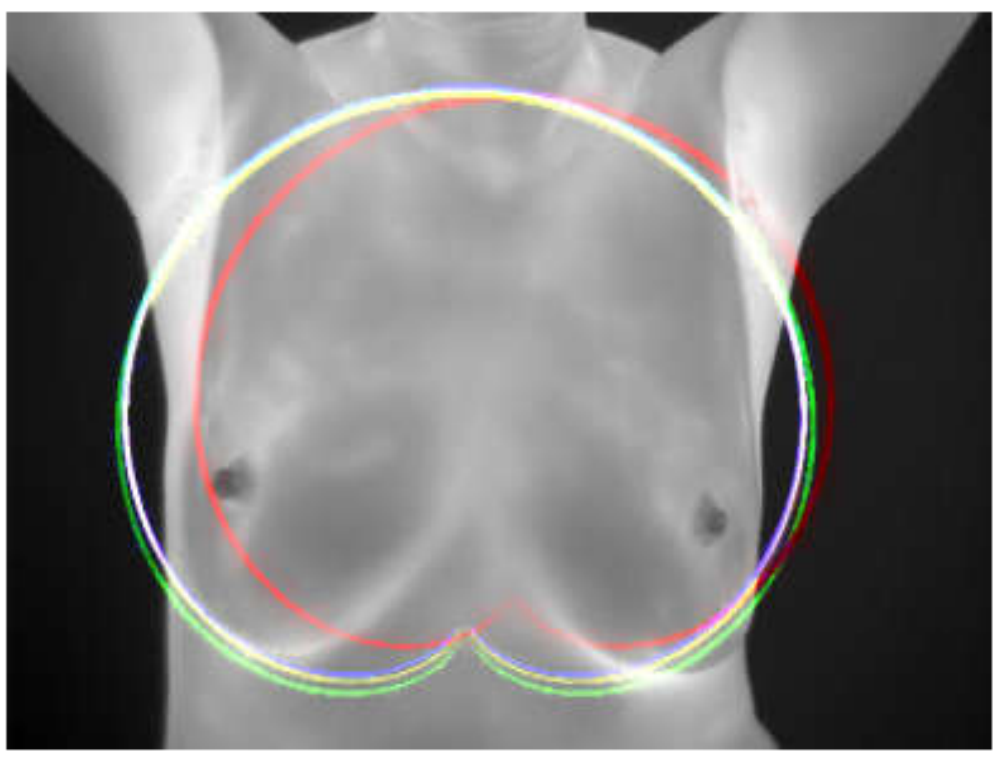

(b) Sick case.

Fig. 5. Cardiod results over different generations.

## V. Discussion

The concept behind this work was to identify individual breasts by 2 independent ellipses (i.e., one for the left breast and the other for the right). However, this approach lead to one ellipse growing over the majority of the area of interest (both breasts and chest), while the other tended to stay small and started to fit itself in one of the arms.

At this point, we started to evaluate two other approaches that consisted of analysing the intersection of the two circles and included this information in the fitness function. This method worked well for healthy patients but for the sick patients it started to fail, mostly because of the discrepant temperature patterns among the left and right breasts in sick patients.

Another major issue was the black metallic marker positioned below the breast during the acquisition of the images. As the black color has a significant negative weight in the fitness function, the circles tended to escape from the region of interest and started to move upwards. This issue still persists even with the use of the cardioid formulation.

We believe that the two ellipses approach is feasible (one ellipse for each breast, and both composing a single individual in the population), however, it requires finetuning. Another possible approach to overcome the black marker is to replace it with random grey information. In this sense, the marker will have little or no effect in the fitness function.

## VI. Conclusion

This work proposes a ROI extraction methodology for thermographic breast images based on genetic algorithms. We exploit the shape of cardioids jointly with gray level data to define a fitness function that is evaluated by a genetic algorithm over time. Our method does not require manual placement of seed points and is therefore entirely automatic, in contrast to other works in the literature [14].

The proposed algorithm was able to provide perfect ROI extraction in 16 out of 58 images and satisfactory results in a further 36 cases. In the case of perfect ROI extraction, we consider a tight and correct ROI extraction with no exclusion of critical breast parts. We consider instances of minor or very little exclusion of critical parts in the breasts as satisfactory results. The remaining 6 cases provided poor results, where the cardiod ended up being displaced over the arm of the patient or another unrelated area.

In terms of time evolution, the GA showed better results after 50 generations, requiring approximately 60 seconds to identify the ROI. Results tend to be better with healthy women as gray level information is more homogeneous. In contrast, sick women (mainly those who have been sick for a long time and with more severe instances of disease) present substantially less homogeneous gray level intensity patterns.

### A. Future Works

The proposed methodology can be further improved in several directions such as using texture information. For instance, information from the application of image processing filters, e.g., anisotropic diffusion, features from the co-occurrence matrix, run-length matrix, sharpening, Laplacian and gradient filters, morphological operations, etc [12], can be accommodated in the fitness function. GA parameters and fitness function can also be further improved and adjusted to produce more accurate results. Similarity computations can also be used [26].

Future work will also consider the effect of applying affine transforms to the cardioid, including rotation. Warps and distortions can also be very efficient when it comes to adjusting the shape of the cardiod to the different types of breasts. The parameters for these transforms can be included in each individual of the GA. Still, regarding GA, new improvements covering different types of operators, evaluation function and selection, as well as new approaches (GA+Fuzzy or GA+RNA) can be explored.